\documentclass[aps,prd,twocolumn,nofootinbib,showpacs]{revtex4}

\newcommand{\be}{\begin{eqnarray}}
\newcommand{\ee}{\end{eqnarray}}
\newcommand{\sfrac}[2]{{\textstyle\frac{#1}{#2}}}
\newcommand{\nn}{\nonumber\\}

\usepackage{bbm}

\newcommand{\m}{\mathbf{m}}
\newcommand{\F}{\mathcal{W}}
\newcommand{\tr}{\mathrm{tr}}

\newcounter{lag}
\newcommand{\lag}{{\stepcounter{lag}{\cal L}_{\arabic{lag}}}}

\newcommand{\B}{\mathbbm{B}}
\newcommand{\M}{\mathbbm{M}}

\usepackage{graphicx}

\newcommand{\confqcd}[1]{~}

\begin{document}

\title{Generalised bottom-up holography and walking technicolour}

\author{Dennis D.~Dietrich}
\affiliation{HEP Center, Institute for Physics and Chemistry, University of
Southern Denmark, Odense, Denmark}
\author{Chris Kouvaris}
\affiliation{The Niels Bohr Institute, Copenhagen, Denmark}

\date{\today}

\begin{abstract}
In extradimensional holographic approaches the flavour symmetry is gauged
in the bulk, that is, treated as a {\it local} symmetry. Imposing such a
local symmetry admits fewer terms coupling the (axial) vectors and
(pseudo)scalars than if a {\it global} symmetry is imposed. The latter is
the case in standard low-energy effective Lagrangians. Here we incorporate
these additional, a priori only globally invariant terms into a holographic 
treatment by means of a
St\"uckelberg completion and alternatively by means of a Legendre
transformation. This work was motivated by our investigations concerning
dynamical electroweak symmetry breaking by walking technicolour and we
apply our findings to these theories.
\end{abstract}

\pacs{
12.60.Nz, 
11.25.Tq, 
12.40.-y, 
12.40.Yx, 
11.15.Tk  
}

\maketitle


\section{Introduction}

Low-energy effective Lagrangians are an extensively used tool when it comes
to the description of the low-energy phenomenology of strongly interacting
theories and chiral symmetry breaking. They are formulated in terms of the
degrees of freedom relevant at low scales. The construction principle for
said Lagrangians derives from the {\it global} symmetries of the physical
system. In the context of strongly interacting theories and chiral symmetry
breaking these are the flavour symmetries.

The same problem setting is addressed by invoking holographic principles
inspired by the AdS/CFT correspondence \cite{Maldacena:1997re}. This
correspondence has originally been conjectured for ${\cal N}=4$
supersymmetry, which is an exactly conformal theory. Hence, using related
methods for non-conformal theories must be seen as extrapolation.
Holographic descriptions of quantum chromodynamics (QCD) yield numbers in
good agreement with experiment \cite{Erlich:2005qh,Da Rold:2005zs} although
in QCD the scale invariance is broken by quantum effects and finite quark
masses. Indeed, holography is currently heavily used as nonperturbative tool
in QCD \cite{Liu:2008tz}. The reason why the holographic principle seems to
be so successful could be the fact that the strong coupling constant appears
to be really a constant below 1 GeV. Simulations on the lattice are also in
favour of the existence of an infrared fixed point
\cite{Furui:2006py,Appelquist:2007hu}.

We are concerned with a very general framework, but a
considerable part of our motivation to investigate these low-energy sectors
comes from dynamical electroweak symmetry breaking through technicolour
theories \cite{TC}. In technicolour models the electroweak symmetry is
broken by chiral symmetry breaking among fermions (techniquarks) in an
additional strongly interacting sector.  (In this respect it is closely
related to quantum chromodynamics.) Walking, that means quasi-conformal
technicolour models \cite{walk} with techniquarks in higher-dimensional
representations \cite{MWT,Dietrich:2006cm,Foadi:2007ue,Belyaev:2008yj} of the technicolour gauge group are consistent
with currently available electroweak precision data. They are thus viable
candidates for breaking the electroweak symmetry dynamically and will be
tested at the Large Hadron Collider (LHC) which is about to become
operational. These walking theories feature by construction a coupling
constant that is constant over a large range of scales (typically, more than
two orders of magnitude). For this reason they meet the criteria for
applying holography in a better way than QCD does
\cite{Hong:2006si,Dietrich:2008ni}. Extra-dimensional holographic frameworks
have even been employed to construct technicolour-like models
\cite{Hirn:2006nt}. Although holography is one of the few tools for
treating strongly coupled field theories and despite its success, the
obtained results should clearly be taken with a grain of salt. Of late, also
lattice simulations for walking technicolour theories are being
carried out \cite{Catterall:2007yx}.

The central issue of the present study is the circumstance that the
holographic principle involves gauging the flavour symmetry in the bulk. On
first sight this turns the flavour symmetry into a {\it local} symmetry.
Imposing a {\it local} symmetry a priori eliminates many terms from the
Lagrangian which would be admissible for a {\it global} symmetry. We here
incorporate the full set of terms admitted by a {\it global} symmetry into a
description involving a {\it local} symmetry by means of a St\"uckelberg
completion, thereby extending previous work on a holographic description of
(minimal) walking technicolour \cite{Dietrich:2008ni}. (In App.~\ref{B} we
provide another equivalent way of achieving this which is based on a Legendre
transformation to antisymmetric tensor fields \cite{DDD}. In the context of effective
Lagrangians a standard procedure is described in \cite{Foadi:2007ue} and
outlined again in App.~\ref{ALTERNATIVE}.) Understandably, when applied to
technicolour the resulting larger number of parameters permits a richer
phenomenology with, for example, an inverted mass hierarchy between vector
and axial vector states in some areas of the parameter space.

The paper is organised as follows. In Sect.~\ref{II} we explain how to
incorporate the additional terms allowed with respect to a global symmetry
into a holographic treatment by means of a St\"uckelberg completion. In
Sect.~\ref{IIA} we derive the spectrum of spin-one degrees of freedom in the
hard-wall model \cite{Polchinski:2000uf,Erlich:2005qh}, in
Sect.~\ref{IIB} for the soft-wall model \cite{Karch:2006pv}. Relative to the
standard holograhic approach the number of parameters is enlarged. In
order to increase the predictive power we discuss possible ways to constrain
the system. First, in Sect.~\ref{IIIA} we revisit the standard holographic
approach where we only had one free parameter \cite{Dietrich:2008ni}.
Second, in Sect.~\ref{IIIB} we impose the first Weinberg sum rule on the level of the lowest
lying resonances, as was done for the four-dimensional treatment of the
corresponding effective Lagrangian in \cite{Foadi:2007ue}. In Sect.~\ref{IV}
we apply our findings to walking technicolour theories---minimal walking
technicolour and beyond. Sect.~\ref{SUMM} summarises the results. In the
appendices we discuss two alternative ways to incorporate the terms
allowed by a global symmetry into a description involving a local symmetry.
App.~\ref{B} is concerned with a Legendre transformation from spin-one to 
spin-two fields. App.~\ref{ALTERNATIVE} discusses another where the
symmetry group is first doubled and subsequently broken spontaneously to the
original symmetry content. 


\section{Holography and effective Lagrangian\label{II}}

Let us consider an effective theory for a global flavour symmetry
$\mathcal{G}$ that is broken to $\mathcal{H}$ by chiral symmetry breaking.
In the case of technicolour, for techniquarks in non-(pseudo)real
representations of the technicolour gauge group this would be the breaking
$SU(N_f)_L\times SU(N_f)_R\rightarrow SU(N_f)_V$, where $N_f$ stands for the
number of techniflavours. For (pseudo)real representations the unbroken
symmetry group $\mathcal{G}$ is enhanced to $SU(2N_f)$. Assuming the
breaking to the maximal diagonal subgroup, it will be broken to $SO(2N_f)$
$[Sp(2N_f)]$ for a real [pseudoreal] representation.

We arrange the left- and right-handed fermion fields $(U,D,\dots)$
of the elementary theory according to,
\be
Q=(U_L,D_L,\dots,-i\sigma^2U_R^*,-i\sigma^2D_R^*,\dots)^T .
\ee
From there we define the bilinear,
\be
M=\epsilon_{\alpha\beta}Q^\alpha\otimes Q^\beta ,
\ee
with contracted spin indices. It contains the scalar and pseudoscalar
degrees of freedom,
\be
M=[\sfrac{1}{2}(\sigma+i\Theta)+\sqrt{2}(i\Pi^a+\tilde\Pi^a)X^a]E .
\ee
$E$ parametrises the condensate which breaks the flavour symmetry and $X^a$
are those generators of ${\cal G}$ which do not commute with $E$.
Per definition, the generators $S^a$ of the residual group ${\cal H}$ commute
with $E$, $ES^aE=-S^{aT}$. (For an explicit realisation of the generators 
and the matrix $E$ in the case of minimal walking technicolour see for
example Ref.~\cite{Foadi:2007ue}.)
Further, we identify the vectorial degrees of freedom,
\be
A_i^{\mu,j}
\sim
Q_i^\alpha\sigma_{\alpha\dot\beta}^\mu\bar{Q}^{\dot\beta,j}
-
\sfrac{1}{4}\delta^j_i
Q^\alpha_k\sigma^\mu_{\alpha\dot\beta}\bar{Q}^{\dot\beta,k} ,
\ee
which populate the entire unbroken group ${\cal G}$, $A_\mu=A^a_\mu T^a$.
With this field content we construct the Lagrangian \footnote{For the use in
technicolour the electroweak gauge bosons would have to be coupled in on
this level as well. We are, however, not considering them here, but
they have been in detail in Refs.~\cite{Foadi:2007ue,DDD}.},
\be
{\cal L}
=
-\sfrac{1}{4g^2_5}F^a_{\mu\nu}F^a_{\kappa\lambda}g^{\mu\kappa}g^{\nu\lambda}
+\sfrac{1}{2} A^a_\mu\m^{ab}A^b_\nu g^{\mu\nu} ,
\label{lag}
\ee
leaving out, for the time being, terms independent of the spin-one fields like kinetic, mass, and
self-interaction terms for $M$. $F^a_{\mu\nu}$ represents the field tensor
of the vectors and is defined as $F^a_{\mu\nu}T^a=F_{\mu\nu}=\partial_\mu
A_\nu-\partial_\nu A_\mu-i[A_\mu,A_\nu]$, where $T^a$ stand for the generators of $\mathcal{G}$. The mass matrix $\m^{ab}$
is symmetric and depends on $M$. The Lagrangian 
is to be invariant under simultaneous {\it global} transformations,
\be
A_\mu &\rightarrow& UA_\mu U^\dagger,
\nn
M&\rightarrow&UMU^T ,
\label{global}
\ee
where $U\in{\cal G}$. Accordingly, the mass term can contain the following
addends
\footnote{The global symmetries also admit a term
$
\sim \tr\{A_\mu[M(i\partial_\nu M^\dagger)-(i\partial_\nu M)
M^\dagger]\}g^{\mu\nu}$, which, however, does not contribute to the
following analysis and is therefore omitted.
}
,
\be
\sfrac{1}{2}A^a_\mu\m^{ab}A^b_\nu g^{\mu\nu}
&=&
[r_1~\tr(A_\mu A_\nu MM^\dagger)
+\label{mass}\\
&&+
r_2~\tr(A_\mu M A_\nu^T M^\dagger)
+
\nn
&&+
s~\tr(A_\mu A_\nu)\tr(MM^\dagger)/(2N_f)]g^{\mu\nu} .
\nonumber
\ee

In a holographic approach, the flavour symmetry is gauged
in the bulk, that is, treated as a gauge symmetry. The action is
postulated to be invariant under {\it local} inhomogeneous transformations,
\be
A^a_\mu T^a=A_\mu\rightarrow U[A_\mu-iU^\dagger(\partial_\mu U)]U^\dagger,
\label{inhom}
\ee
of the vector field (and simultaneous transformations of $M$). The kinetic
term ($FF$) for the spin-one particles is already invariant under these
inhomogenous transformation. To the contrary, they do not
leave invariant the general mass term (\ref{mass}). 
The only a priori
invariant combination is a gauged kinematic term for the (pseudo)scalars
$\sim\tr[(D_\mu M)(D_\nu M)^\dagger]g^{\mu\nu}$, with the covariant
derivative $D_\mu M:=\partial_\mu M-iA_\mu M-iMA_\mu^T$. It contains only a
special combination of the addends comprised in Eq.~(\ref{mass}). (See also Sect.~\ref{CONSTRAINT}.)

For the more general term, the invariance under inhomogeneous transformations
(\ref{inhom}) can be restored through the introduction of non-Abelian 
St\"uckelberg degrees of freedom $\Phi=e^{-i\theta^a T^a}$.
The mass term (\ref{mass}) then turns into,
\be
&&A^a_\mu\m^{ab}A^b_\nu g^{\mu\nu}
\rightarrow
(A_\mu^\Phi)^a
\m^{ab}
(A_\nu^\Phi)^b
g^{\mu\nu} ,
\ee
where $A^\Phi_\mu=A_\mu-i\Phi(\partial_\mu\Phi)^\dagger$. The field $\Phi$
transforms like $\Phi\rightarrow U\Phi$ under local transformations and
consequently, $A_\mu^\Phi\rightarrow UA_\mu^\Phi U^\dagger$.
Also here a radial, that is, Higgs-like degree of freedom ensures
perturbative renormalisability. Said degree of freedom together with the
field $\Phi$ is related to the field $N$ in
Ref.~\cite{Foadi:2007ue}. (See also App.~\ref{ALTERNATIVE}.)

Once we proceed with the holographic analysis, we have to decide what to do
with the St\"uckelberg degrees of freedom. In a general gauge they will obey
their proper equation of motion which has to be solved simultaneously with
the equations of motion for $A_\mu$ and $M$. In unitary gauge we set
$\Phi\equiv 1$. In that case, we do not have to keep track of these degrees
of freedom, but we cannot enforce additionally the frequently used $A_5\equiv 0$
gauge. We shall pursue this approach in what follows. Another alternative is to
translate from the language of spin-one fields $A_\mu$ to spin-two field
$\tilde B_{\mu\nu}$ as demonstrated in App.~\ref{B}. There the St\"uckelberg
fields are automatically absorbed in the spin-two fields.

Subsummarizing, the terms of the five-dimensional action in unitary gauge 
relevant for the following calculations are,
\be
S:=\int d^5x \sqrt{g} ({\cal L}+{\cal L}_M),
\label{action}
\ee
where,
\be
{\cal L}_M
:=
\tr[(\partial_\mu M)(\partial_\nu M)^\dagger] g^{\mu\nu}
-
m^2_5\tr(MM^\dagger),
\ee
and with ${\cal L}$ given by Eq.~(\ref{lag}).
The metric is anti de Sitter,
\be
ds^2=z^{-2}(-dz^2+dx^2).
\ee
The fifth coordinate, $z$, is interpreted as inverse energy scale. The
coupling $g_5$ is fixed through matching to perturbative calculations,
\be
g^2_5=12\pi^2/d_\mathrm{R},
\ee
where $d_R$ stands for the dimension of the technicolour gauge group with respect to which the fermions transform.

According to the action (\ref{action}) the expectation value $M_0 E$, $E$=constant, of the field $M$ obeys the equation of motion,
\be
(z^3\partial_zz^{-3}\partial_z-z^{-2}m^2_5)M_0=0.
\label{eomm0}
\ee
The characteristic equation for the dimension $d$ of the scalar operator is obtained from the ansatz $M_0\sim z^d$,
\be
m^2_5=d(d-4).
\label{char}
\ee
In the quasi-conformal case $d=2$ and the general solution of (\ref{eomm0}) can be written as,
\be
M_0=c_1 z^2+c_Wz^2\ln(z/\epsilon),
\ee
where $c_1$ and $c_W$ depend on the boundary conditions. In general, in the ultraviolet  ($z=\epsilon$) $M_0$ is proportional to the techniquark mass matrix. Hence, in the chiral limit, $c_1=0$. $c_W$ is treated below.

In four-dimensional momentum ($q$) space the equation of motion for the transverse part $A_\perp^a$ of the spin-one field is found to be,
\be
[(z\partial_z z^{-1}\partial_z +q^2)\delta^{ab}-g^2_5z^{-2}\mathbf{m}^{ab}]A^b_\perp=0.
\ee
It differs from the minimal holographic approach \cite{Dietrich:2008ni,Erlich:2005qh,Karch:2006pv,Hong:2006si} by the appearance of the mass matrix $\mathbf{m}^{ab}$. Diagonalisation in flavour space leads to,
\be
{}[(z\partial_z z^{-1}\partial_z +q^2)-g^2_5z^{-2}M_0^2\lambda^2_\F]\F=0,
\label{spinone}
\ee
where hereinafter $\F$ always stands as representative for the vector
$\mathcal{V}$ and the axial vector $\mathcal{A}$ eigensolution,
respectively. The corresponding eigenvalues of $\m^{ab}/M_0^2$ are given by,
\be
\lambda^2_\mathcal{V}&=&r_1-r_2+s,\\
\lambda^2_\mathcal{A}&=&r_1+r_2+s.
\ee

The approximation $M_0=c_Wz^2\ln(z/\epsilon)\approx c_W z^2\ln(z_m/\epsilon)$
allows us to determine analytic expressions for the eigensolutions $\F$. The
thus introduced error is only slight: In the hard-wall model $z$ lives on
the interval $[\epsilon,z_m]$. The approximation deviates most for small
values of $z$, of the order of $\epsilon$, and goes to zero for large values
of $z$ around $z_m$. In Eq.~(\ref{spinone}) the term involving $M_0$ is
negligible in the ultraviolet. Hence, where the approximation is not that
accurate it does not play a role. To the contrary, in the infrared, where it
counts, the approximation is accurate. With a soft wall $z$ has no sharp
bound, but is still cut off gradually by the potential. As we shall argue,
there is no need to specify a value for $z_m$ because different choices
amount merely to redifinitions of constants. We thus write,
\be
M_0\approx Cz^2/g_5,
\label{approx}
\ee
which in the hard-wall model would correspond to $C\approx
g_5c_W\ln(z_m/\epsilon)$.

For determining the pion decay constant, we have hence to solve the
differential equation for the axial part [(\ref{spinone}) with
$\F\rightarrow\mathcal{A}$] with $q^2=0$ and for the boundary conditions,
\be
\partial_z\mathcal{P}(z_m)=0\mathrm{~~~and~~~}\mathcal{P}(0)=1.
\label{bcp}
\ee 
$f_\pi$ is obtained from the solution by,
\be
g^2_5f_\pi^2=-\lim_{\epsilon\rightarrow 0}\partial_z\mathcal{P}(\epsilon)/\epsilon,
\label{fpi}
\ee
which for $\partial_z\mathcal{P}(0)=0$ turns into $-\partial^2_z\mathcal{P}(0)$.


\subsection{Hard-wall model\label{IIA}}

In the so-called hard-wall approach \cite{Polchinski:2000uf,Erlich:2005qh}
$z$ takes values on the interval $[\epsilon,z_m]$. $z_m$ corresponds to
the position of the infrared boundary and $\epsilon$ to that of the
ultraviolet. In the context of quasi-conformal theories which (almost)
exhibit a conformal behaviour over a range of scales, the aforementioned
interval can be identified with this range \cite{Hong:2006si}. In
phenomenologically viable technicolour models $\epsilon\ll z_m$ and we can
take the limit $\epsilon\rightarrow 0$, which is smooth. Thus, the boundary
conditions for the spin-one fields are given by,
\be
\F(0)=0=\partial_z\F(z_m).
\ee
For $\F(0)=0$ the solutions of the equations of motion (\ref{spinone}) with
$M_0$ from (\ref{approx}) are given by,
\be
\F\sim z^2e^{-C_\F z^2/2}M(1-\sfrac{q^2}{4C_\F},2,C_\F z^2),
\ee
where $C_\F=C\lambda_\F$ and $M(\dots)$ represents a Kummer function. The
boundary condition at $z=z_m$ implies,
\be
(2C^2_\F z_m^2-M_\F^2)M(1-\sfrac{M_\F^2}{4C_\F},2,C_\F z_m^2)
&+&
\nn
+
(4C_\F+M_\F^2)M(-\sfrac{M_\F^2}{4C_\F},2,C_\F z_m^2)
&=&
0.
\ee
It yields the values for the masses for the vectorial and axial states, but can only be solved numerically. Imposing normalisation conditions on the solutions,
\be
\int_0^{z_m}\frac{dz}{z}\F^2=1,
\ee
the decay constants can be extracted according to,
\be
g_5F_\F=\partial^2_z\F(0),
\label{defdechard}
\ee
which also requires numerical calculations.

The solution for $\mathcal{P}$ which obeys the boundary conditions
(\ref{bcp}) reads,
\be
\mathcal{P}
=
\cosh(C_\mathcal{A}z^2/2)-\tanh(C_\mathcal{A}z_m^2/2)\sinh(C_\mathcal{A}z^2/2),
\nn
\ee
and extracting the pion decay constant $f_\pi$ according to Eq.~(\ref{fpi}) yields,
\be
g^2_5f_\pi^2=C_\mathcal{A}\tanh(C_\mathcal{A}z_m^2/2).
\ee


\subsection{Soft-wall model\label{IIB}}

The soft-wall model \cite{Karch:2006pv} is based on the incorporation of an additional dilaton field $\phi$, such that the action becomes,
\be
S_s:=\int d^5x\sqrt{g}e^{-\phi}(\mathcal{L}+\mathcal{L}_M).
\ee
From the requirement that the mass spectrum show Regge behaviour, that is a
linear spacing of the squared masses, it follows that the dilaton background
has to behave like $cz^2$, $c$=constant, for large $z$. Instead of the
previous potential well, we have now to deal with a harmonic oscillator. The
infrared boundary conditions is substituted by the normalisability of the
solution on $\mathbbm{R}^+$.

First we have to determine the behaviour of the condensate $M_{0,s}$ in the
new setting. It satisfies the equation of motion,
\be
(z^3e^{+cz^2}\partial_z z^{-3}e^{-cz^2}\partial_z-z^{-2}m^2_5)M_{0,s}=0.
\ee 
The characteristic equation (\ref{char}) holds approximately for $c^2z^2\ll
m^2_5$. Hence, by means of an identification in the ultraviolet we can once
more set $m^2_5=-4$. The previous differential equation has the solution,
\be
g_5M_{0,s}=C z^2 e^{+cz^2}.
\ee
$M_{0,s}$ grows exponentially with $z$. This
hints towards the presence of an instability which must ultimately be cured
by non-linear terms in the differential equation \cite{Karch:2006pv}. They
would come from potential terms for the (pseudo)scalars not addressed here.
Therefore, the above solution is appropriate for small values of $z$ and we
can set
\be
g_5M_{0,s}\mapsto Cz^2.
\label{M0s}
\ee
On one hand, one can interpret the last step as practical way of regularising
the aforementioned instability. There are actually nonlinear terms which would
modify the equation of motion precisely in such a way that Eq.~(\ref{M0s}) is an
exact solution. On the other hand, one can see the previous step as one 
introducing an $O(cz^2)$ approximation. For that latter case, the effect of the 
thus introduced error will be assessed below.

The equations of motion for the spin-one fields read,
\be
[(ze^{+cz^2}\partial_z z^{-1}e^{-cz^2}\partial_z+q^2)-g^2_5z^{-2}M_0^2m^2_\F]\F_s=0.
\nn
\ee
With the substitution,
\be
\F_s
=:
\tilde\F_s
e^{+cz^2/2},
\ee
we find
\be
[(z\partial_zz^{-1}\partial_z+q^2)-g^2_5z^{-2}M_0^2\lambda^2_\F-c^2z^2]
\tilde\F_s=0.
\ee
Taking into account the ultraviolet boundary condition right away, for $M_{0,s}$ from (\ref{M0s}) this equation is solved by
\be
\tilde\F_s
\sim
z^2e^{-c_\F z^2/2}M(1-\sfrac{M_\F^2}{4c_\F},2,c_\F z^2),
\ee
where $c^2_\F:=c^2+C^2_\F$. This function is normalisable only if the Kummer function truncates into a polynomial, which it does for $M^2_\F=4nc_\F$, $n\in\mathbbm{N}$. For $n=1$ the normalisation condition,
\be
\int_0^\infty\frac{dz}{z}\tilde\F^2=1,
\ee
selects the unique solution
\be
\F=\sqrt{2}c_\F z^2.
\ee
This leads to the decay constant,
\be
g_5F_\F=\partial_z^2\F(0)=2\sqrt{2}c_\F=M_\F^2/\sqrt{2}.
\label{softdecay}
\ee

In order to determine the pion decay constant $f_\pi$ we have to solve the
differential equation,
\be
[ze^{+cz^2}\partial_z z^{-1}e^{-cz^2}\partial_z-g^2_5z^{-2}M_0^2\lambda^2_\F]\mathcal{P}_s=0,
\ee
which after the substitution
\be
\mathcal{P}_s
=:
\tilde\mathcal{P}_s
e^{+cz^2/2},
\ee
becomes
\be
[z\partial_zz^{-1}\partial_z-g^2_5z^{-2}M_0^2\lambda^2_\F-c^2z^2]
\tilde\mathcal{P}_s=0.
\ee
With $M_{0,s}$ from (\ref{M0s}) and for the boundary conditions
(\ref{bcp}) this equation has the solution,
\be
\tilde\mathcal{P}_s=e^{-c_\mathcal{A}z^2/2}.
\ee
Equivalently,
\be
\mathcal{P}_s=e^{-(c_\mathcal{A}-c)z^2/2},
\ee
which through the relation (\ref{fpi}) yields,
\be
g^2_5f_\pi^2=c_\mathcal{A}-c.
\label{fpic}
\ee
Together with $c>0$ and $M_a^2=4c_\mathcal{A}$ this relation leads directly
to a lower bound for the mass of the lightest axial vector, 
$M_a^2\ge4g_5^2f_\pi^2$.

If we interpret Eq.~(\ref{M0s}) as $O(cz^2)$ approximation instead of as regularisation we have to assess its range of applicability. To this end we calculate the shift $\delta q^2_F$ of the masses $M^2_\F$ originating from the difference $\delta U$ between the exact and the approximate potential. The normalised wave functions were given by
\be
\tilde\F_s=\sqrt{2}c_\F z^2e^{-c_\F z^2/2},
\ee
and the difference between the potentials,
\be
\delta U=C_\F^2 z^2(e^{+cz^2}-1).
\ee
This leads to the shift,
\be
\delta q^2_\F
&=&
\int_0^\infty\frac{dz}{z}~\tilde\F_s^2~\delta U
=
\nn
&=&
2C_\F^2 c_\F^2[(c_\F-c)^{-3}-c_\F^{-3}].
\ee
It should be smaller than the actual masses $M_\F^2=4c_\F$. Hence,
\be
x^2_\F\sqrt{1+x_\F^2}[(\sqrt{1+x_\F^2}-1)^{-3}-(1+x_\F^2)^{-3/2}]\ll 2,
\nn
\ee
where $x_\F=C_\F/c$. The left-hand side is divergent for small $x_\F$ and
approaches zero for large $x_\F$. Hence, we need $C_\F\gg c$. Exploiting the
axial part of this statement in combination with Eq.~(\ref{fpic}) leads to,
\be
x_\mathcal{A}=\sqrt{(1+y)^2-y^2}/y,
\ee
where $y:=c/(g^2_5f_\pi^2)$. Thus, we need large $x_\mathcal{A}$ and hence
small $y$ or $c\ll g^2_5f_\pi^2$. Results from an iterative study of an
analogous set of equations in \cite{Dietrich:2008ni} showed that for
$c=g^2_5f_\pi^2$ the deviation was still only ten percent. Through
Eq.~(\ref{fpic}) this implies 
$
M_a^2/(4g^2_5f_\pi^2)\lesssim2$.


\section{Constraining the parameter space\label{CONSTRAINT}}

The incorporation of the mass term (\ref{mass}) increases the number of
parameters for our present analysis effectively by one. One could regard the
coupling strength of the vectorial eigenstates to the condensate as this
parameter. It was equal to zero in the non-generalised approach.


\subsection{Non-generalised holography\label{IIIA}}

In the standard holographic approach \cite{Erlich:2005qh,Dietrich:2008ni} the 
masses of the spin-one fields arise from the kinetic term of the $M$ field,
$
g^{\mu\nu}\tr[(D_\mu M)(D_\nu M)^\dagger]
$.
It contains,
\be
g^{\mu\nu}\tr[(A_\mu M+MA_\mu^T)(A_\nu M+MA_\nu^T)^\dagger],
\ee
which corresponds to $r_1=2=r_2$ and $s=0$. Hence
$\lambda_\mathcal{V}^2=0$, that is, the vectorial degrees of freedom are not
coupled to the condensate. The unconstrained analysis only depends on the
ratio of $r_1+s$ and $r_2$. The overall magnitude of these parameters can be
absorbed in a rescaling. Thus, the above three conditions only fix one
degree of freedom.


\subsection{Weinberg sum rules\label{IIIB}}

Another way to fix a degree of freedom is to assume that the first
Weinberg sum rule \cite{Weinberg:1967kj}---which holds for asymptotically
free gauge field theories---be satisfied already at the level
of the lowest resonances \cite{Foadi:2007ue}:


\subsubsection{Soft-wall model}
Let us start the discussion with the oblique $S$ parameter
\cite{Peskin:1990zt} as calculated from the lowest resonances. 
(From this point onward, we will use the subscript $a$ for the lightest axial
vector resonance and $\rho$ for the lightest vector resonance.) It can be
interpreted as the zeroth sum rule,
\be
\frac{S}{4\pi}=\frac{F_\rho^2}{M_\rho^4}-\frac{F_a^2}{M_a^4}.
\label{S}
\ee
Due to Eq.~(\ref{softdecay}) it is exactly equal to zero because,
\be
\frac{F_\rho^2}{M_\rho^4}=\frac{1}{2g^2_5}=\frac{F_a^2}{M_a^4}.
\ee
From the first Weinberg sum rule,
\be
f_\pi^2
=
\frac{F_\rho^2}{M_\rho^2}-\frac{F_a^2}{M_a^2},
\ee
we find accordingly,
\be
2g^2_5f_\pi^2
=
M_\rho^2-M_a^2,
\ee
implying immediately that the vector must be heavier than the axial vector.
The lightest vector mass, $M_\rho^\mathrm{min}=\sqrt{6}g_5f_\pi$ is realised for
the lightest axial mass, $M_a^\mathrm{min}=2g_5f_\pi$. According to 
Eq.~(\ref{softdecay}) the associated minimal decay constant is given by,
$(F_\rho^\mathrm{min})^{1/2}=18^{1/4}g^{1/2}_5f_\pi$.
(See also Tab.~\ref{Tab1}.)

In quasi-conformal theories the second Weinberg sum rule receives
corrections \cite{Appelquist:1998xf},
\be
8\pi^2\alpha f_\pi^4/d_\mathrm{R}
=
F_\rho^2-F_a^2,
\ee
with the parameter $\alpha>0$. (Otherwise, $F_\rho^2-F_a^2=0$.) After imposing the first sum rule we find for this parameter,
\be
\frac{\alpha}{3}
=
1+\frac{M_a^2}{g^2_5f_\pi^2}.
\ee


\subsubsection{Hard-wall model}

In the hard-wall approach the expressions for the various masses and decay
constants must be evaluated numerically and we present sum-rule
constrained results together with the values for the soft-wall approach in
the figures.


\section{Walking technicolour\label{IV}}

\begin{table}[t]
\begin{tabular}{ccccccccc}
\hline\hline
$N_c$&rep&$d_\mathrm{R}$&$N_f$&$N_f^g$&
$M_{a,\mathrm{soft}}^\mathrm{min}$&$(F_{a}^\mathrm{min})^{1/2}$&
$M_{\rho,\mathrm{soft}}^\mathrm{min}$&$(F_{\rho}^\mathrm{min})^{1/2}$\\
\hline
2&F&2&7&6&2.2&0.66&2.7&0.81\\
2&F&2&7&2&3.8&1.15&4.6&1.41\\
2&G&3&2&2&3.1&1.04&3.8&1.27\\
3&F&3&11&2&3.1&1.04&3.8&1.27\\
3&S&6&2&2&2.2&0.87&2.7&1.07\\
3&G&8&2&2&1.9&0.81&2.3&0.99\\
4&F&4&15&2&2.7&0.97&3.3&1.18\\
4&S&10&2&2&1.7&0.77&2.1&0.94\\
4&A&6&8&2&2.2&0.87&2.7&1.07\\
4&G&15&2&2&1.4&0.69&1.7&0.85\\
5&F&5&19&2&2.4&0.91&2.9&1.12\\
5&A&10&6&2&1.7&0.77&2.1&0.94\\
6&F&6&23&2&2.2&0.87&2.7&1.07\\
\hline
&&&&&TeV&TeV&TeV&TeV\\
\hline\hline
\end{tabular}
\caption{
Various walking technicolour models from Tab.~III in
\cite{Dietrich:2006cm}. Minimal (axial) vector meson mass
$M_{a/\rho,\mathrm{soft}}^\mathrm{min}$ and square root of the minimal (axial) vector
decay constant $(F_{a/\rho}^\mathrm{min})^{1/2}$ for various walking technicolour
models characterised by the representation R (F=fundamental, G=adjoint, 
S=two-index symmetric, A=two-index antisymmetric) of the technicolour gauge group
under which the techniquarks transform and the number of (gauged) techniflavours $N_f$ ($N_f^g$).
}
\label{Tab1}
\end{table}

Ref.~\cite{Dietrich:2006cm} lists viable candidates for dynamical
electroweak symmetry breaking by technicolour theories. They can display a
sufficiently large amount of walking and are consistent with
electroweak precision data. For the present analysis they are characterised
by the number of techniflavours $N_f$ and the dimension $d_\mathrm{R}$ of the
representation $\mathrm{R}$ of the gauge group under which the techniquarks
transform. The prime candidate, minimal walking technicolour, features two
techniquarks in the adjoint representation of $SU(2)$. The adjoint
representation is real and therefore the symmetry breaking pattern is
$SU(4)\rightarrow SO(4)$.

Primarily here, we present results for minimal walking technicolour. Different numbers
of flavours or dimensions lead to rescalings of the masses and decay
constants. Relative to minimal walking technicolour the masses are
multiplied by $(3/d_R)^{1/2}$ and $(2/N_f)^{1/2}$. {\it For this scaling},
$N_f$ is given by the number of techniflavours gauged under the electroweak
gauge group. This must be an even number to have complete doublets. It has
proven to be advantageous to include more than two techniquarks to be close
to conformality, while only gauging two under the electroweak to avoid the
bounds from precision data. For these {\it partially gauged}
\cite{MWT,Dietrich:2006cm} technicolour models, the number of gauged
flavours is given under $N_f^g$ in Tab.~\ref{Tab1}.
The decay constants of the spin-one mesons have a different scaling
behaviour because of the additional factor of $g_5$ in
Eqs.~(\ref{defdechard}) and (\ref{softdecay}). Relative to their values in
minimal walking technicolour the $F^{1/2}$ scale like $(3/d_R)^{1/4}$ and
$(2/N_f)^{1/2}$. 
%
\begin{figure}[t]
\includegraphics[width=8.7cm]{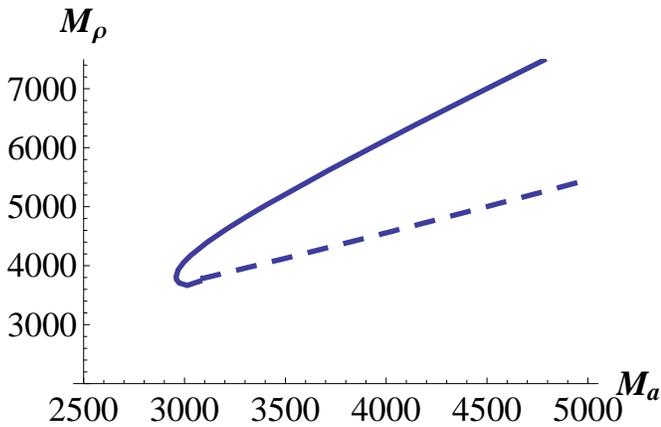}
\caption{
Minimal walking technicolour constrained by the first Weinberg sum-rule: The
mass $M_\rho$ of the first vector meson as function of the mass $M_a$ of the
first axial vector meson in the hard-wall model (solid) and in the soft-wall
model (dashed). At the low-mass end in the hard-wall model, different values
of $M_\rho$ for a given $M_a$ arise for different values of the underlying
parameters $C_A$ and $z_m$. All numbers are in GeV.} \label{Fig1}
\end{figure}
%
In order to give a quantitative idea about this aformentioned scaling the
minimal values for the masses in the soft-wall approach and the associated
minimal values for the decay constants are given in the last four columns of
Tab.~\ref{Tab1}.

Minimal walking technicolour features an enhanced symmetry breaking pattern
$SU(4)\rightarrow SO(4)$ relative to non-(pseudo)real representations and
the most basic $SU(2)_L\times SU(2)_R\rightarrow SU(2)_V$ pattern known from
two-flavour QCD. Of the models listed in Tab.~\ref{Tab1} only those with
techniquarks in the two-index symmetric representation of either $SU(3)$
or $SU(4)$, feature this simple pattern. In that case only mesonic
fields are contained in the action (\ref{action}). There are exactly three
pions which are absorbed to become the longitudinal degrees of freedom of
the electroweak gauge bosons.

The models with techniquarks in the fundamental representation of $SU(N>2)$
have the symmetry breaking pattern $SU(N_f)_L\times SU(N_f)_R\rightarrow
SU(N_f)_V$---always assuming the breaking to the minimal diagonal
subgroup---which leads to a larger field content due to the larger number
of flavours. The same holds for the model based on the two-index antisymmetric
representation of $SU(5)$. The symmetry does not mix left and right fields. Consequently,
we have only mesonic states among the (pseudo)scalars and (axial) vectors.
There are, however, more than the three pions which are absorbed to become
the longitudinal degrees of freedom of the electroweak gauge bosons.
The decomposition into vector and axial vector eigenstates is straightforward, 
$2\mathcal{V}=A_L+A_R$ and $2\mathcal{A}=A_L-A_R$.

\begin{figure}[t]
\includegraphics[width=8.7cm]{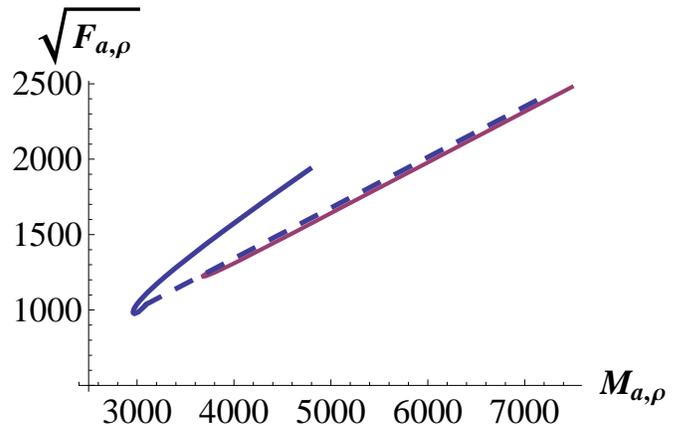}
\caption{Minimal walking-technicolour constrained by the first Weinberg 
sum-rule: The square root of the decay constant of the first vector meson 
$F_\rho^{1/2}$ as function of its mass $M_\rho$ (hard-wall: thin solid, 
soft-wall: dashed) and the square root of the decay constant of the first 
axial vector meson $F_a^{1/2}$ as function of its mass $M_a$ (hard-wall: 
thick solid, soft-wall: dashed), 
respectively. At the low-mass end in the hard-wall model, different values of 
$M_\rho$ for a given $M_a$ arise for different values of the underlying 
parameters $C_A$ and $z_m$. All numbers are in GeV.} 
\label{Fig2}
\end{figure}

The adjoint representation is always real, and all corresponding models with
different numbers of colours feature two flavours, including minimal walking
technicolour. Here the symmetry mixes left and right fields. Therefore,
fields with non-zero technibaryon number contribute to the action
(\ref{action}) \footnote{Technibaryons in technicolour theories with
techniquarks in (pseudo)real representations of the technicolour gauge group
are also studied as components of dark matter \cite{DM}.}.

The other models, seven flavours of the fundamental representation of
$SU(2)$ as well as eight flavours of the two-index antisymmetric
representation of $SU(4)$ have enhanced symmetries, as they contain more
than two flavours, and because they are based on pseudoreal and real
representations, respectively. Thus part of the flavour symmetries involve
mixing left- and right-handed fields and part of the spin-zero and spin-one
states carry technibaryon number. Nevertheless, the spin-one particles are
always either vector or axial vector mass eigenstates when described within
the present framework. The detailed decomposition which involves $SU(14)$
or $SU(16)$ generators, respectively, is not performed here.


\subsection{Weinberg sum-rules}

The results from standard holography have been discussed in great detail in
Ref.~\cite{Dietrich:2008ni}. We here turn immediately to the results from
the analysis constrained by the first Weinberg sum-rule. After imposing this
constraint one free parameter remains. We choose it to be the mass $M_a$ of
the lightest axial state.

The vector mass $M_\rho$ as function of the axial vector mass $M_a$ is shown 
in Fig.~\ref{Fig1}. For small axial vector masses $M_a$, the mass $M_\rho$ of
the first vector state coincides in the two scenarios, hard- and soft-wall,
respectively. Both masses are bounded from below. The axial vector mass is
bounded from below due to its relation to the physical value of the pion
decay constant $f_\pi$. In the soft wall model,
\be
M_a>2g_5f_\pi.
\ee
The vector mass in the soft-wall model is a monotonously rising function of
the axial vector mass and, hence, the minimum is reached for the minimal
axial-vector mass,
\be
M_\rho>\sqrt{6}g_5f_\pi.
\ee

\begin{figure}[t]
\includegraphics[width=8.7cm]{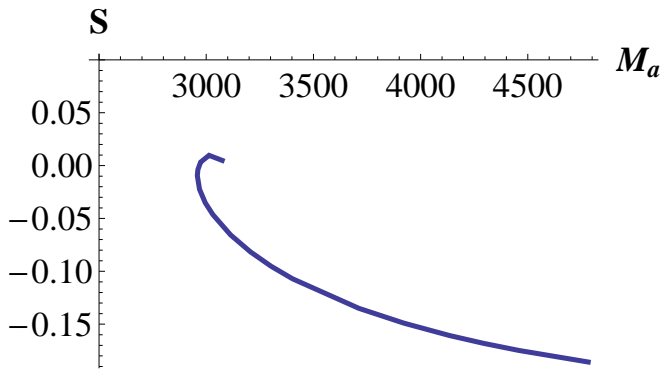}
\caption{
Minimal walking technicolour constrained by the first Weinberg sum-rule:
Oblique $S$ parameter calculated from the lowest (axial) vector resonances
as function of the lowest axial vector mass $M_a$ in the hard-wall model. 
$M_a$ is in GeV.}
\label{Fig3}
\end{figure}

In the hard-wall model, one can go slightly below the bounds known
analytically for the soft-wall model. In the hard-wall approach in the
low-mass region there are also two values for the vector mass for a given
value of the axial vector mass. There is also a small
region of values of $M_\rho$ for which two values of $M_a$ can be found.
The two different solutions correspond to different values of the underlying
parameters $C_\mathcal{A}$ and $z_m$. 
In the soft-wall model towards larger masses, the vector mass slowly
approaches the axial vector mass from above.
This is not seen in the hard-wall model.

Fig.~\ref{Fig2} displays the decay constants as functions of the mass of the
corresponding state. In the soft-wall approach the curves of the vector and
the axial vector coincide. The vectorial curve begins at higher values of
the mass though. We have always $M_\F^2/F_\F=\sqrt{2}g_5$. At low masses the hard-wall results coincide
with that of the soft wall. In that region the hard infrared bound is so far
away from the ultraviolet bound that the states essentially do not feel it.
The hard-wall model curve for the vector stays always very close (but
slightly below) the soft-wall curve. This is indicative of the fact that the
vector couples so strongly to the condensate that it also does (almost) not
feel the hard wall. For increasing mass the decay constant of the axial
vector state departs to larger values from the soft-wall result and
consequently also the hard-wall vectorial result. It is more strongly
influenced by the infrared boundary condition.

\begin{figure}[t]
\includegraphics[width=8.7cm]{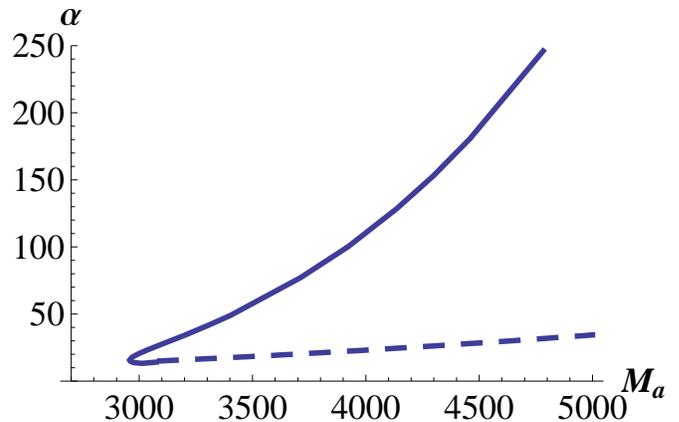}
\caption{Minimal walking-technicolour constrained by the first Weinberg sum-rule:
$\alpha$-parameter from the modified second Weinberg sum-rule as function of the lowest axial vector mass $M_a$ in the hard-wall (solid) and the soft-wall model (dashed), respectively.
$M_a$ is in GeV.} \label{Fig4}
\end{figure}

The contribution to the oblique $S$ parameter from the lowest resonances
computed according to Eq.~(\ref{S}) for the hard-wall model is displayed in 
Fig.~\ref{Fig3} as a function of the axial mass $M_a$. At low mass it starts
out with small positive values $<0.02$ and for increasing mass decreases 
towards negative values of the order $-0.1$. 
For the soft-wall model the oblique $S$ parameter is identical to zero.
These values are consistent with the experimental values reported in
\cite{Amsler:2008zz}.

\begin{figure}[t]
\includegraphics[width=8.7cm]{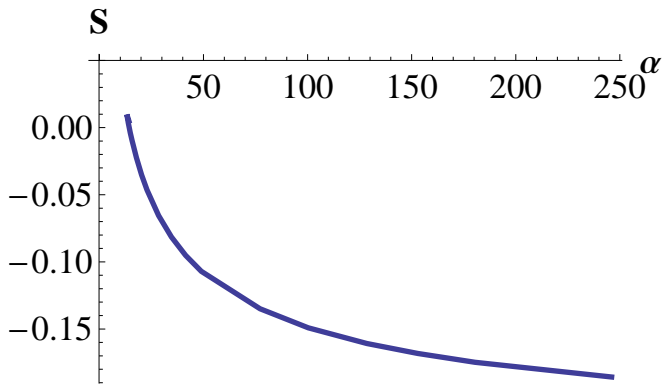}
\caption{
Minimal walking-technicolour constrained by the first Weinberg sum-rule:
Oblique $S$ parameter calculated from the lowest (axial) vector resonances
as function of the $\alpha$-parameter from the modified second Weinberg
sum-rule in the hard-wall model.} \label{Fig5}
\end{figure}

A reduction of the $S$ parameter in quasi-conformal theories relative to
its perturbative value is expected due to the modification of the second
Weinberg sum rule \cite{Appelquist:1998xf}. A reduction to negative values, 
however, has not been anticipated: The parameter $\alpha$ characterising the
modification is expected to be of order one which reduces $S$, but does not
suffice to change its sign \cite{Appelquist:1998xf}. Here, however, the parameter $\alpha$ grows
beyond order unity. (See Fig.~\ref{Fig4}.) In view of this fact in the
hard-wall model, decreasing $S$ is still related to growing $\alpha$. (See
Fig.~\ref{Fig5}). In the soft-wall model the $\alpha$ parameter grows much
more slowly with increasing axial mass $M_a$, but the $S$ parameter is
constant and equal to zero.


\section{Summary\label{SUMM}}

We have introduced a generalisation of the standard bottom-up holographic method
\cite{Erlich:2005qh,Hong:2006si,Dietrich:2008ni,Polchinski:2000uf,Karch:2006pv}.
The generalisation consists of admitting all addends in a mass term for the spin-one
states which are invariant under a {\it global} symmetry transformation. All these
terms would be present in the non-extradimensional treatment of an effective
Lagrangian. In the holographic method, however, the flavour symmetry is gauged in
the bulk and is thus {\it local}. A priori, a local symmetry admits fewer terms. 
In order to be able to incorporate the additional terms, we carry out a St\"uckelberg completion, in the main body of the paper.
In the appendix two alternative methods are presented; one based on a Legendre 
transformation to spin-two fields \cite{DDD} and one based on a doubling and 
subsequent spontaneous breaking of the symmetry \cite{Foadi:2007ue}.

After laying out the framework, we determine the spectrum of low-lying spin-one 
states for the case of quasi-conformal theories in the hard- \cite{Polchinski:2000uf,Erlich:2005qh}
and the soft-wall approach \cite{Karch:2006pv}. In the generalised approach there
is effectively one more parameter than in the standard version. In order to increase
predictive power, a way to constrain anew the parameter space is to impose the
first Weinberg sum rule on the level of the lowest resonances. The first Weinberg 
sum rule holds in general for asymptotically free gauge field theories. That it is 
(approximately) satisfied on the level of the first resonances is an assumption. 
We discuss the results for the thus constrained system and apply it to walking 
technicolour theories. Also and especially there the imposition of sensible 
constraints is essential: In a generic effective Lagrangian approach it is crucial
to link the numerously arising parameters in the effective Lagrangian to 
parameters in the elementary theory. For QCD one can conveniently rely on 
experimental data. Data for physics beyond the standard model is much sparser. Here
theoretically motivated postulates must be used.

Naturally, when the same four-dimensional effective Lagrangian is treated with or 
without extradimensions, the results differ in most of the larger parameter space 
of the non-extradimensional treatment. As a concrete example, in the 
non-extradimensional treatment presented in \cite{Foadi:2007ue}, one 
parameter/degree of freedom was eliminated instead by setting the 
perturbative oblique $S$ parameter in the elementary 
equal to the $S$ parameter in the effective theory. To the contrary, in the present 
holographic treatment the $S$ parameter comes as output, with a reduction relative
to its perturbative value which is generally expected \cite{Appelquist:1998xf} in 
the presence of walking dynamics.

There are results from the holographic approach independent of the additional
constraints, that is, whether standard holography, the first Weinberg sum rule or
no constraint is imposed. On one hand, there is the scaling behaviour of the masses and 
decay constants with the dimension $d_\mathrm{R}$ of the representation of the 
technicolour gauge group under which the techniquarks transform and the number $N_f$ 
of techniflavours gauged under the electroweak interactions. The masses scale 
like $d_\mathrm{R}^{-1/2}$ and $N_f^{-1/2}$; the decay constants like 
$d_\mathrm{R}^{-1/4}$ and $N_f^{-1/2}$. On the other hand, the minimal required 
mass of the axial vector is always linked to the physical value of the pion decay 
constant $f_\pi$, $M_a>4\pi[6/(N_fd_\mathrm{R})]^{1/2}\Lambda_\mathrm{ew}$. Further, in the 
soft-wall model the ratio between the squared mass of a spin-one state and its 
decay constant is always $M^2_\F/F_\F=\sqrt{2}g_5$.

Other features differ between the differently constrained holographic scenarios. 
While in standard holography the axial vector mass is always larger than the 
vector mass, the imposition of the first Weinberg sum rule in the extended scheme
leads to an inversion of the mass hierarchy. Without any constraint, the 
generalised holography can accomodate both orderings of the mass. (After all, 
standard holography is only a special case of the generalised version.)

In walking theories, an oblique $S$ parameter which is reduced with respect
to its perturbative value is expected \cite{Appelquist:1998xf}, not,
however, one which turns negative. If this feature is enforced it selects
the soft-wall model where $S=0$ or the hard-wall model at small axial-vector
masses.


\section*{Acknowledgments}

The authors would like to thank
Roshan Foadi,
Mads T.~Frandsen,
Deog-Ki Hong,
Matti J\"arvinen, 
and
Francesco Sannino
for helpful and inspiring discussions.
The work of DDD was supported by the Danish Natural Science Research Council. 
The work of CK was supported by the Marie Curie Fellowship under contract MEIF-CT-2006-039211.


\appendix

\section{Legendre transformation of the spin-one sector\label{B}}

In this appendix we discuss another way for promoting a global to a local
symmetry. It is based on a Legendre transformation from spin-one to spin-two
fields which is generally possible for non-Abelian gauge field theories.
(See for example \cite{DDD} and references therein.) Applying the
Legendre transformation to the effective theory can be interpreted in two
ways. On one hand, we can say that the low-energy theory has already been
elevated to a gauge theory by introducing St\"uckelberg degrees of freedom
(see above) which then properly allows to carry out the Legendre
transformation. The practical approach, on the other hand, is to carry out
the Legendre transformation directly for the effective theory, as one would
do for a gauge theory. The outcome is the same and for this reason, the
second interpretation can be seen as working prescription for promoting the
effective theory to a gauge theory.


\subsection{Gauge theory}

As announced just above, we will use a specific Legendre transformation
which is generally feasible for a gauge theory to promote an effective
theory to a gauge theory. In the following we will always present the
Lagrangian densities for the sake of brevity, but with the implicit
understanding that an action or even a partition function is constructed
from them. For the sake of clarity, let us first look at a prototypical
massive Yang--Mills gauge theory,
\be
\lag
=
-\sfrac{1}{4g^2_5}F^a_{\mu\nu}F^a_{\kappa\lambda}g^{\mu\kappa}g^{\nu\lambda}
+\sfrac{m^2}{2}A^a_\mu A^a_\nu g^{\mu\nu} .
\ee
This Lagrangian is not gauge invariant under inhomogeneous transformations
(\ref{inhom}) of the vector field. The Lagrangian must therefore again be 
supplemented by non-Abelian St\"uckelberg fields,
\be
\lag
=
-\sfrac{1}{4g^2_5}F^a_{\mu\nu}F^a_{\kappa\lambda}g^{\mu\kappa}g^{\nu\lambda}
+m^2\tr[(D_\mu\Phi)^\dagger(D_\nu\Phi)]g^{\mu\nu} .
\nn
\ee
An economic way to incorporate an antisymmetric tensor field
$\tilde B^a_{\mu\nu}$---and finally to carry out the aforesaid Legendre
transformation---is to introduce a Gaussian term into the Lagrangian,
\be
\lag
&=&
[-\sfrac{1}{4g^2_5}F^a_{\mu\nu}F^a_{\kappa\lambda}
+\sfrac{g^2_5}{4}\tilde B^a_{\mu\nu}\tilde B^a_{\kappa\lambda}]
g^{\mu\kappa}g^{\nu\lambda}
+
\nn
&&+
m^2\tr[(D_\mu\Phi)^\dagger(D_\nu\Phi)]g^{\mu\nu} ,
\ee
and subsequently to shift the $\tilde B^a_{\mu\nu}$ field by a multiple 
$\sfrac{1}{g^2_5}F^a_{\mu\nu}$ of the field tensor. This leads to,
\be
\lag
&=&
[\sfrac{g^2_5}{4}\tilde B^a_{\mu\nu}\tilde B^a_{\kappa\lambda}
-\sfrac{1}{2}\tilde B^a_{\mu\nu}F^a_{\kappa\lambda}]
g^{\mu\kappa}g^{\nu\lambda}
+
\nn
&&+
m^2\tr[(D_\mu\Phi)^\dagger(D_\nu\Phi)]g^{\mu\nu} .
\ee
The St\"uckelberg fields can now be absorbed in the antisymmetric tensor
fields: A gauge transformation of the Yang--Mills potential yields,
\be
\lag
&=&
[\sfrac{g^2_5}{2}
\tr(\Phi\tilde B_{\mu\nu}\Phi^\dagger\Phi\tilde B_{\kappa\lambda}\Phi^\dagger)
-\tr(\Phi\tilde B_{\mu\nu}\Phi^\dagger F_{\kappa\lambda})]
\times
\nn
&&\times
g^{\mu\kappa}g^{\nu\lambda}
+
\sfrac{m^2}{2}A^a_\mu A^a_\nu g^{\mu\nu} ,
\ee
where we made use of $\Phi^\dagger\Phi\equiv 1$. In a partition function the
functional integral for $\tilde B^a_{\mu\nu}$ runs already over all possible
configurations. Integrating independently over all configurations of $\Phi$
merely corresponds to carrying out the former integral several times. Hence,
the St\"uckelberg fields can be omitted and the corresponding integration
factors out from a partition function. (The previous Lagrangian corresponds to a non-Abelian
non-linear sigma model, which when quantised is not perturbatively
renormalisable. This can be circumvented by introducing a position-dependent
dynamical mass, that is a Higgs degree of freedom. The lastmentioned step
does not interfere with the previous gauge symmetry arguments.) Therefore,
we can continue working with the unitary gauge expression,
\be
\lag
=
[\sfrac{g^2_5}{4}\tilde B^a_{\mu\nu}\tilde B^a_{\kappa\lambda}
-\sfrac{1}{2}\tilde B^a_{\mu\nu}F^a_{\kappa\lambda}]
g^{\mu\kappa}g^{\nu\lambda}
+
\sfrac{m^2}{2}A^a_\mu A^a_\nu g^{\mu\nu} ,
\nn
\ee
from which we would like next to eliminate the gauge potential. As the
Lagrangian is of at most second order in the Yang--Mills field, the
resulting Lagrangian up to fluctuation terms reads,
\be
\lag
&=&
\sfrac{g^2_5}{4}\tilde B^a_{\mu\nu}\tilde B^a_{\kappa\lambda}
g^{\mu\kappa}g^{\nu\lambda}
-\label{lagb}\\
&&-\sfrac{1}{2}
(\sfrac{1}{\sqrt{g}}\partial_\kappa\sqrt{g}\tilde B^{a\kappa\mu})
(\M^{-1})^{ab}_{\mu\nu}
(\sfrac{1}{\sqrt{g}}\partial_\lambda\sqrt{g}\tilde B^{b\lambda\nu}) ,
\nonumber
\ee
where
$\M^{bc}_{\mu\nu}=\B^{bc}_{\mu\nu}-m^2\delta^{bc}g_{\mu\nu}$, the
inverse of which exists, in general, for more than two space-time
dimensions, and $\B^{bc}_{\mu\nu}:=f^{abc}\tilde B^a_{\mu\nu}$. Quantum
fluctuations of the Yang--Mills potential induce a term which on the
Lagrangian level is proportional to $\sfrac{1}{2}\ln\det\M$.  Lorentz
indices are raised by the metric $g^{\mu\nu}$. $\sqrt{g}$ stands for the
square root of the determinant of the metric. The whole section has entirely 
been written in terms of the two-form field $\tilde B^a_{\mu\nu}$, so that the
expressions become outwardly independent of the number of space-time
dimensions. This field is the dual of a $(d-2)$-form field
$B^a_{\mu_1,\dots,\mu_{d-2}}$. 

For more details on these and related issues in the context of gauge field
theories as such, the reader is referred to Refs.~\cite{DDD}.


\subsection{Effective theory}

We now proceed to the effective theory described by the Lagrangian
(\ref{lag}) for which we have carried out the St\"uckelberg completion
above. Again, as in the previous subsection, when it comes to the
reformulation in terms of antisymmetric tensor fields, the St\"uckelberg
fields will be absorbed. Thus, repeating those steps leads to the equivalent
of Eq.~(\ref{lagb}),
\be
\lag
&=&
\sfrac{g^2_5}{4}\tilde B^a_{\mu\nu}\tilde B^a_{\kappa\lambda}
g^{\mu\kappa}g^{\nu\lambda}
-\\
&&-\sfrac{1}{2}
(\sfrac{1}{\sqrt{g}}\partial_\kappa\sqrt{g}\tilde B^{a\kappa\mu})
(\mathbbm{m}^{-1})^{ab}_{\mu\nu}
(\sfrac{1}{\sqrt{g}}\partial_\lambda\sqrt{g}\tilde B^{b\lambda\nu}) ,
\nonumber
\ee
where $\mathbbm{m}^{bc}_{\mu\nu}:=\mathbbm{B}^{bc}_{\mu\nu}-\mathbf{m}^{bc}g_{\mu\nu}$.
The term induced by quantum fluctuations is proportional to
$\sfrac{1}{2}\ln\det\mathbbm{m}$. 

As a first application, let us look at the effective potential in
$B^a_{\mu\nu}$-field formulation. Notably, as opposed to the standard
representation it is manifestly gauge invariant.
\be
V_\mathrm{eff}
=
\sfrac{g^2_5}{4}\tilde B^a_{\mu\nu}\tilde B^a_{\kappa\lambda}
g^{\mu\kappa}g^{\nu\lambda}
-\sfrac{\mu}{2}\ln\det\mathbbm{m}.
\ee
It contains contributions from quantum fluctuations encoded in the
determinant term. $\mu$ is a dynamically induced scale which can be
interpreted as the inverse volume of a lattice cell. It can be fixed by
a renormalisation condition. 
The characteristic equation for the minimum of the potential reads,
\be
g_5^2\tilde B^a_{\mu\nu}=\mu(\mathbbm{m}^{-1})^{bc}_{\mu\nu}f^{abc}.
\ee
If $\mathbf{m}^{bc}$ in $\mathbbm{m}^{bc}_{\mu\nu}$ dominates over 
$\mathbbm{B}^{bc}_{\mu\nu}$,
the previous saddle-point condition becomes approximately,
\be
[\delta^{ag}-g^2_5\mu(\mathbf{m}^{-1})^{db}f^{gbc}(\mathbf{m}^{-1})^{ec}f^{dea}]
\tilde B^a_{\mu\nu}
\approx
0.
\ee
In general, this equation is solved by $\tilde B^a_{\mu\nu}=0$, in which
case the assumption that $\mathbf{m}^{bc}$ dominate is trivially justified.
Nontrivial solutions arise if the determinant of the expression inside
square bracktes vanishes. 
For $\mathbbm{B}^{bc}_{\mu\nu}$ dominating over $\mathbf{m}^{bc}$, the lowest order calculation corresponds to the massless case, 
\be
g_5^2\tilde B^a_{\mu\nu}\approx \mu(\mathbbm{B}^{-1})^{bc}_{\mu\nu}f^{abc}.
\ee
This equation does evidently not admit the solution $\tilde B^a_{\mu\nu}=0$.
It has been studied in Euclidean space \cite{Schaden:1989pz} which requires
numerical tools.


\section{Local symmetries in effective Lagrangians\label{ALTERNATIVE}}

We recapitulate here the recipe for introducing a {\it local} symmetry laid
out in Ref.~\cite{Foadi:2007ue} and combine it with the
holographic treatment. The (pseudo)scalars $M$ transform globally under the
representation $\mathrm{R}$ of the group $\mathcal{G}$. The spin-one field $A_\mu^a$
transform under the adjoint representation of a copy $\mathcal{G}^\prime$ of
the original flavour group $\mathcal{G}$, under which it transforms as
singlet. The field $M$ is a singlet under the group $\mathcal{G}^\prime$. In
order to connect the two fields in the two different groups a scalar field
$N$ is introduced which transforms under the fundamental representation of
$\mathcal{G}$ and the antifundamental of $\mathcal{G}^\prime$. $A_\mu$ is
promoted to a gauge field over $\mathcal{G}^\prime$ such that the covariant
derivative of $N$ reads,
\be
D_\mu N=\partial_\mu N-iNA_\mu .
\ee
In order to recover the original symmetry content, $N$ is made to acquire a
vacuum expectation value $\langle N^i_j\rangle\sim\delta^i_j$ such that the
semisimple embedding group $\mathcal{G}\times\mathcal{G}^\prime$ is broken
to its diagonal subgroup $\mathcal{G}_V$, which is again a copy of
$\mathcal{G}$. The corresponding condensate must be much stronger than the
one associated to the chiral/electroweak symmetry breaking in the actual
theory.  The field $P_\mu^a$ defined through,
\be
2i\tr(N^\dagger N)P_\mu
=
[(D_\mu N)N^\dagger-ND_\mu N^\dagger]d_\mathrm{F},
\ee
when evaluated on the vacuum expectation value of the field $N$ reproduces
$A^a_\mu$, $\langle P_\mu^a\rangle=A_\mu^a$. Finally, the kinetic terms are
given by,
\be
\mathcal{L}_\mathrm{kin}
&=&
-\sfrac{1}{2g^2_5}\tr(F_{\mu\nu}F_{\kappa\lambda})g^{\mu\kappa}g^{\nu\lambda}
+
\nn
&&+
\tr[(D_\mu N)(D_\nu N)^\dagger]g^{\mu\nu}
+
\nn
&&+
\tr[(\partial_\mu M)(\partial_\nu M)^\dagger]g^{\mu\nu}.
\label{lkin}
\ee
The $P^a_\mu$ and $M$ fields are linked by replacing $A^a_\mu$ by
$P^a_\mu$ in the mass term (\ref{mass}). 

In the holographic treatment we pursued above we had first determined the
vacuum expectation value for the field $M$ and then evaluated the action for
the spin-one fields on this expectation value. The vacuum expectation value
for $N$ can be incorporated in the same way. Hence, the present treatment
amounts to evaluating the action for the spin-one fields on the expectation
values of the spin-zero fields $M$ and $N$. 
Per definition, $P^a_\mu$ evaluated on the expectation value of $N$
coincides with $A^a_\mu$. Hence, all occurences of $P^a_\mu$ can be replaced
by $A^a_\mu$ and from this point onward we join the analysis laid out in the
main body of the paper.

The kinetic term for the $N$ fields in Eq.~(\ref{lkin}) when evaluated on
the expectation value of $N$ contributes a hard mass term for the spin-one
fields, $\sim\tr(A_\mu A_\nu^\dagger)g^{\mu\nu}\tr(NN^\dagger)$. As
has already been said above, the important point concerning the condensate of 
$N$ is that it is much stronger than that of $M$. If we put the $N$ condensate
proportional to the $M$ condensate the inclusion of the $N$ condensate amounts 
effectively only to a
redefinition of the parameter $s$. As $r_1$ and $r_2$ are completely
arbitrary and $s$ appears exclusively in the combination $r_1+s$, this has no
influence on the subsequent analysis. A different condensate function for
$N$ can be introduced with the same right and effect as a different dilaton
background.


\end{document}